\documentclass[prl,aps,twocolumn,superscriptaddress,dvipsnames,longbibliography]{revtex4}
\pdfoutput=1
\usepackage{graphicx}
\usepackage{amsmath,amssymb,amsthm,amsfonts,mathptmx}
\usepackage{color}
\usepackage{algorithm}
\floatname{algorithm}{Protocol}
\usepackage[labelsep=period]{caption}
\usepackage{ftnxtra}

\usepackage{times}

\theoremstyle{plain}

\theoremstyle{definition}

\newcommand{\bra}[1]{\langle#1|}
\newcommand{\ket}[1]{|#1\rangle}

\def \BQP {\textbf{BQP}}
\def \QMA {\textbf{QMA}}
\def \NP {\textbf{NP}}
\def \PP {\textbf{PP}}

\begin{document}
\title{Post hoc verification of quantum computation}
\author{Joseph F. Fitzsimons}\email{joseph_fitzsimons@sutd.edu.sg}
\affiliation{Singapore University of Technology and Design, 8 Somapah Road, Singapore 487372}
\affiliation{Centre for Quantum Technologies, National University of Singapore, 3 Science Drive 2, Singapore 117543}

\author{Michal Hajdu\v{s}ek}
\affiliation{Singapore University of Technology and Design, 8 Somapah Road, Singapore 487372}
\affiliation{Centre for Quantum Technologies, National University of Singapore, 3 Science Drive 2, Singapore 117543}

\begin{abstract}
With recent progress on experimental quantum information processing, an important question has arisen as to whether it is possible to verify arbitrary computation performed on a quantum processor. A number of protocols have been proposed to achieve this goal, however all are interactive in nature, requiring that the computation be performed in an interactive manner with back and forth communication between the verifier and one or more provers. Here we propose two methods for verifying quantum computation in a non-interactive manner based on recent progress in the understanding of the local Hamiltonian problem. Provided that the provers compute certain witnesses for the computation, this allows the result of a quantum computation to be verified after the fact, a property not seen in current verification protocols.\end{abstract}

\maketitle

Quantum computers offer the potential to dramatically increase our ability to efficiently solve otherwise intractable computational problems, spanning a range of applications, from computational number theory \cite{Shor1994} to the simulation of physical systems \cite{lloyd1996universal}. While for most of its history the experimental side of the field has been restricted to performing experiments on systems with very few degrees of freedom, which could be simulated using conventional computers with modest effort, recent advances have begun to push against this bound \cite{trotzky2012probing}. As the complexity of controlled quantum systems has grown, it has begun to exceed our ability to convincingly answer questions as to whether observed behaviour is consistent with quantum mechanics. This question has already arisen in relation to several recent experiments, leading to sometimes heated debate \cite{boixo2013experimental,smolin2013classical,wang2013comment,boixo2014evidence,ronnow2014defining,shin2014quantum,vinci2014distinguishing,shin2014comment,albash2015reexamining}.

Fortunately, recent years have also seen the emergence of interactive methods for verifying quantum dynamics, in the form of protocols for verified quantum computation \cite{Aharonov:2008,Fitzsimons:2012,Barz:2013,Reichardt:2013,Mckague:2013,Morimae2014}. Such protocols can be divided broadly into two classes. The first class are protocols which use a small, well characterized, quantum device to verify a computation carried out on a larger quantum system \cite{Aharonov:2008,Fitzsimons:2012,Barz:2013,Morimae2014,hayashi2015verifiable}. Protocols in this class generally exhibit low overhead, but this comes at the cost of an assumption that there is no malicious conspiracy involving both the verifier's device and the quantum system being probed. The second class of protocols make use of queries to multiple non-communicating quantum provers which share entanglement to verify the computation using self-testing techniques \cite{Reichardt:2013,reichardt2013classical2,Mckague:2013}. Such protocols offer security conditioned under the assumption that the provers do not communicate during the protocols. This type of security comes at the price of astronomical overhead, and enforcing a ban on communication between provers can become problematic in multi-round protocols where space-like separation becomes infeasible. Recently attempts have been made to unify the two approaches, making use of self-testing protocols to replace the quantum operations of the verifier in verifiable blind computation protocols \cite{gheorghiu2015robustness,hajduvsek2015device}, which has shown some success in lowering the overheads associated with multi-prover schemes.

Despite their differences, several common features remain between all known protocols from both classes. Current protocols require continuous interaction during the computation in order to verify correctness \footnote{In the case of schemes derived from the blind quantum computing approach of Morimae and Fujii \cite{Morimae:2013}, this amounts to the prover pacing the rate at which they send qubits to the verifier, even though communication is only in one direction.}. This fundamentally links the task of verification to the specific implementation of the computation, making it impossible to verify the correctness of a computation after the fact. Furthermore, all known verification schemes, whether intentional in the design or not, exhibit some form of blind quantum computation \cite{Broadbent:2009,Barz:2012}, allowing them to hide the computation being performed from the processor performing the computation.

Here we introduce a new approach to the problem of verifying quantum computation, based on recent progress on short interactive proofs of the local Hamiltonian problem \cite{fitzsimons2015multiprover,ji2015classical}. This gives rise to a pair of verification protocols that can be used to verify a quantum computation with only a single round of communication. This verification can occur arbitrarily long after the computation has been completed, provided that a suitable witness state has been prepared. Furthermore, these post hoc verification protocols do not inherently exhibit blindness, and hence their existence serves to introduce a separation between the notions of blindness and verifiability.

The class of decision problems answerable by a quantum computer in polynomial time (\BQP) is contained within a larger class known as \QMA. This larger class corresponds to the class of problems for which the solution can be verified by a quantum computer in polynomial time given a suitable quantum state to act as a witness \cite{kitaev1999quantum,aharonov2002quantum}. \QMA~can be though of as the quantum analogue of the classical complexity class \NP. Of the problems known to be complete for \QMA, meaning that any problem within \QMA~can be reduced to an instance of that problem with modest computational overhead, perhaps the most studied is the problem of deciding whether the ground state energy of some local Hamiltonian is bounded below some threshold \cite{kempe20033,kempe2006complexity}. Specifically, given constants $a$ and $b$, with $a<b$, and a Hamiltonian $H=\sum_{i=1}^m H_i$, where each $H_i$ acts non-trivially on at most $k$ qubits, with the promise that for constants $a<b$ either $\bra{\psi}H\ket{\psi}\leq am$ for some $\ket{\psi}$ or $\bra{\psi}H\ket{\psi}\geq bm$ for all $\ket{\psi}$, the local Hamiltonian problem is to decide between these two cases.

The fact that \BQP~is contained in \QMA~does not immediately imply a non-trivial method to verify a quantum computation, since several conditions must be met in order for a verification protocol to be useful:
\begin{itemize}
\item Not every computation we may wish to perform with a quantum computer is a decision problem, and so in order to exploit membership of \QMA~in constructing a verification procedure, it is first necessary to cast the verification as a decision problem.
\item The definition of \QMA~allows for the verification to require polynomial-sized quantum circuits, and hence it is not clear \textit{a priori} that verification can be any more efficient than the original computation. 
\item It must be possible to efficiently generate a witness state for a given quantum computation. In general, participation as a prover in interactive proofs may require significantly greater computational power than that required to decide the particular statement to be proved.
\end{itemize}
We now show that each of these criteria can be satisfied by casting the problem of verifying the output of a quantum circuit as an instance of the local Hamiltonian problem.

Our first step is to formalize the verification of a quantum circuit as a decision problem. There are a number of ways in which this problem can be posed. A strong definition of this problem was introduced in \cite{Dunjko2014} based on trace distance between the output state at the end of the protocol and the ideal output of the computation to be verified. This definition is effectively impossible to decide retrospectively given a finite number of runs of the computation, due to the lack of any observable to compare probability distributions. Instead, we consider a formulation motivated by operational concerns, which sidesteps the problem of comparing two probability distributions:
\textit{Given a quantum circuit $\mathcal{C}$ composed of initial input in the state $\ket{0}^{\otimes{n}}$ followed by a polynomial number of one- and two-qubit gates chosen from some standard gate set, let $M$ be a string obtained by sampling the output of $\mathcal{C}$ in the computational basis. Given a string $S$, and the promise that either the probability $p_{S}$ with which $S=M$ is at least $1-\delta$ or at most $1-\delta-\gamma$ for some positive $\gamma$, the probability verification problem is to decide which of these is the case.}

\begin{figure}
\center
\includegraphics[width=0.6\columnwidth]{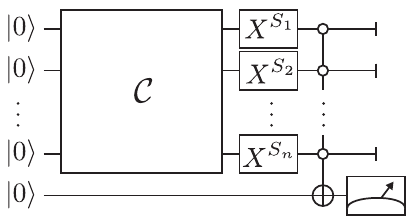}
\caption{A quantum circuit for verifying that $S$ is a possible output of computation $\mathcal{C}$. The measured qubit is in state $\ket{1}$ with probability $p_S$.\label{fig:simple}}
\end{figure}

\begin{figure}
\center
\includegraphics[width=0.7\columnwidth]{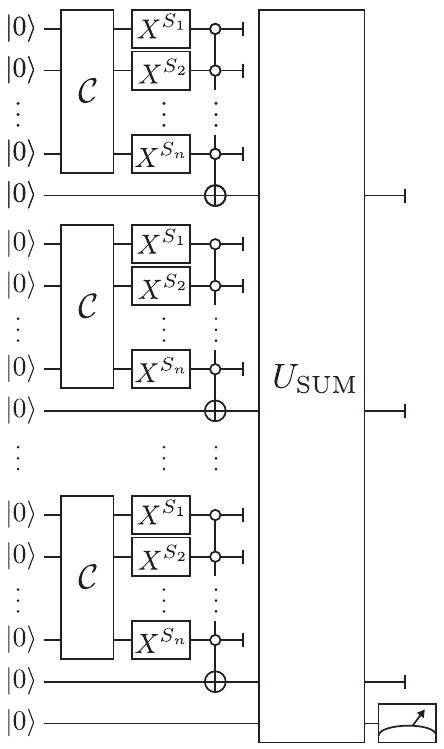}
\caption{A quantum circuit for verifying that $S$ is a possible output of computation $\mathcal{C}$ with amplified probability of success. $U_\text{SUM}$ performs an $X$ gate on the final qubit conditioned on at least a fraction $1-\delta-\frac{\gamma}{2}$ of the other qubits being in state $\ket{0}$. From Hoeffding's inequality \cite{hoeffding1963probability}, it follows that when $p_{S}\geq 1-\delta$ that the probability $\tilde{p}_S$ that output qubit is measured to be in state $\ket{1}$ with probability at least $1-\exp(-\frac{N\gamma^2}{2})$, where $N$ is the number of times $\mathcal{C}$ is evaluated. However, when $p_S\leq \epsilon$ then $\tilde{p}_S$ is at most $\exp(-\frac{N\gamma^2}{2})$.\label{fig:Ncopies}}
\end{figure}

Intuitively, this problem captures the task of deciding whether $S$ is indeed a likely outcome of the chosen computation described by $\mathcal{C}$ or not. We will restrict attention to the case where $\gamma$ is bounded from below by some inverse polynomial in the number of qubits strictly greater than zero, since in the case where the gap can be arbitrarily small this problem becomes \PP-hard \cite{postselection}. The problem of verifying that $S$ was produced according to some particular probability distribution is removed. A simple quantum circuit for this decision problem is shown in Fig. \ref{fig:simple}. Measuring the output qubit in the computational basis results in $\ket{1}$ with probability precisely equal to $p_{S}$. Provided that $\gamma$ is at most polynomially small, this decision problem is then contained within \BQP. This procedure can be extended to amplify the probability of accepting only when $p_S$ is above $1-\delta$ as shown in Fig. \ref{fig:Ncopies}. We shall refer to this latter circuit as the verification circuit $\mathcal{V}_{(\mathcal{C},S)}^N$ for $\mathcal{C}$ for output $S$.

We now turn attention to the task of casting the verification problem described above as an instance of the local Hamiltonian problem. As part of the proof that the local Hamiltonian problem is \textbf{QMA}-complete, Kitaev showed how one can encode a quantum computation as the ground state of a 5-local Hamiltonian \cite{kitaev1999quantum}. Specifically, he showed that a 5-local Hamiltonian could be constructed such that its ground state has energy below some threshold if and only if a chosen quantum circuit, from which Hamiltonian is constructed, results in a specific output qubit being in state $\ket{1}$ with high probability. Here we make use of a subsequent improvement by Kempe, Kitaev and Regev which reduces the  required locality for Hamiltonian to 2-local \cite{KempeKitaevRegev}. Given a quantum circuit $\mathcal{C}'$ composed of $T$ one- and two-qubit gates acting on some initial state $\ket{\psi(0)}$ which can be taken to be $\ket{0}^{\otimes n}$, with a designated output qubit used to decide whether the circuit accepts or rejects, this is accomplished by making use of a \textit{clock-state},
\begin{equation*}
\ket{\eta} = \sum_{t=0}^T \ket{1}^{\otimes t} \ket{0}^{\otimes T-t} \otimes \ket{\psi(t)},
\end{equation*}
where $\ket{\psi(t)}$ is the state of the computation after the first $t$ gates have been applied. The state $\ket{\eta}$ is simply a superposition of states of a clock register comprising the first $T$ qubits of the system concatenated with the state of the logical computation after the number of steps indicated by the clock. In the original 5-local construction, the guiding intuition is that Hamiltonian terms of fixed locality can be used to separately verify that the clock state is of an appropriate form by ensuring that a $\ket{0}$ never appears to the immediate left of a $\ket{1}$, ensuring that neighbouring values of the clock register differ by the corresponding gate on the computational register, and ensuring that input bits and output bits are in the desired state when the clock begins with $\ket{0}$ or ends with $\ket{1}$ respectively.

The 2-local construction is somewhat more involved, and makes use of perturbation theory to reduce the locality of the operators required to verify that the state $\ket{\eta}$ is of the required form. Specifically, the Hamiltonian corresponding to $\mathcal{C}'$ is given by
\begin{equation*}
H_{\mathcal{C}'} = H_\text{out} + J_\text{in} H_\text{in} + J_2 H_\text{prop2} + J_1 H_\text{prop1} + J_\text{clock} H_\text{clock},
\end{equation*}
where $J_\text{in} \ll J_2 \ll J_1 \ll J_\text{clock} \leq \text{poly}(N)$. The definitions of the individual Hamiltonians $H_\text{out}$, $H_\text{in}$, $H_\text{prop2}$, $H_\text{prop1}$ and $H_\text{clock}$ are relatively complicated, and the reader is referred to Ref. \cite{KempeKitaevRegev} for a full definition, however their role is straightforward. $H_\text{out}$ is used to ensure that the output qubit is in state $\ket{1}$ when the right most qubit of the clock register is in state $\ket{1}$, by applying an energy penalty when the clock register ends in $\ket{1}$ but the output qubit is $\ket{0}$. $H_\text{in}$ similarly ensures that the computational register is in state $\ket{0}^{\otimes N}$ when the clock register begins with $\ket{0}$. $H_\text{clock}$ ensures that the clock register is restricted to states of the form $\ket{1}^{\otimes t} \ket{0}^{T-t}$ for $0 \leq t\leq T$. Finally, $H_\text{prop1}$ and $H_\text{prop2}$ ensure that the computational register differs by an appropriate logic gate for branches of the wavefunction corresponding to neighbouring clock states. Importantly for the sake of our verification approach, provided that the gate set is well chosen, each Hamiltonian contains products of only Pauli $X$ and $Z$ operators. Furthermore, the ground state of the Hamiltonian can be prepared efficiently by simply preparing an ancillary register in a superposition of states $\ket{0}$ to $\ket{T}$ and then performing the computation in a controlled manner up to the gate indicated by this register.

Kempe, Kitaev and Regev proved that if $\mathcal{C}'$ accepts with probability more than $1-\delta$, then $H_{\mathcal{C}'}$ has an eigenvalue smaller than $\delta$, whereas if $\mathcal{C}'$ accepts with probability less than $\epsilon$, then all eigenvalues of $H_{\mathcal{C}'}$ are larger than $\frac{1}{2} - \epsilon$. As such, if circuit encoded in the Hamiltonian is taken to be the verification circuit $\mathcal{V}_{(\mathcal{C},S)}^N$ for a particular quantum computation $\mathcal{C}$ and output $S$, as depicted in Fig. \ref{fig:Ncopies}, then deciding the local Hamiltonian problem for $H_{\mathcal{V}_{(\mathcal{C},S)}^N}$, with constants $a=\frac{\exp(-N\gamma^2/2)}{m}$ and $b=\frac{1 - 2\exp(-N\gamma^2/2)}{2 m}$ is equivalent to verifying the original computation.

The above construction alone does not suffice to provide a means to verify a quantum computation without making use of a comparably powerful device, since the task of estimating the energy of the witness state $\ket{\eta}$ may well be comparable in complexity to independently implementing the circuit to be evaluated. However, if the number of provers is expanded, the complexity of verifying the ground state drops. Specifically, it was shown recently in Ref. \cite{fitzsimons2015multiprover}, that there existed a one-round multi-prover interactive proof for the $k$-local Hamiltonian problem where the verifier sends a classical message of length logarithmic in the number of local terms in the Hamiltonian to each of each of 5 provers each of whom respond with at most $k$ qubits. The interactive proof takes the form of a simple game, which can be won by honest provers if they share the ground state to the Hamiltonian encoded in a specific five-qubit quantum error correction code, with each prover holding one of the physical qubits comprising each logical qubit \footnote{The number of provers can trivially be reduced to four by instead making use of an error-detection code.}.

The game proceeds as follows for a $k$-local Hamiltonian $H = \sum_{i=1}^m H_i$ on $n$ qubits. The verifier decides with equal probability either to test $H$ or to test the encoding. If the verifier decides to test $H$, then they choose a $j$ uniformly at random between $1$ and $m$, and ask each prover for their share of the qubits corresponding to the set $S_j$ of qubits acted upon by $H_j$. The verifier then decodes the returned qubits, and performs a measurement on the decoded qubits with POVM elements $H_j$ and $\mathbb{I}-H_j$, rejecting if the outcome corresponds to $H_j$ and accepting otherwise. If instead the verifier decides to test the encoding, they then choose uniformly randomly between each of two subtests. In the first, they choose an $i$ uniformly at random between $1$ and $n$, and ask each prover for their share of qubit $i$. They then accept if the returned qubits lie within the code space, and reject if not. In the second subtest, they choose both an $i$ uniformly at random between $1$ and $n$, and an $S$ as a uniformly random tuple of three distinct integers between $1$ and $n$ such that $S$ contains $i$. They then ask a random prover for their share of qubits indexed by $S$, and the remaining provers for their share of qubit $i$. As in the other subtest, the verifier accepts if the shares of qubit $i$ lie within the code-space, and reject otherwise.

Intuitively, the second test is used to ensure that the provers must always respond with distinct qubits associated with the indices requested, so that their response when a particular qubit $q$ is requested does not depend on which subset of qubits containing $q$ they have been asked for. The first test then simply ensures that these (now fixed) qubits correspond to a valid witness for the specific instance of the local Hamiltonian problem being tested. It was proved in \cite{fitzsimons2015multiprover} that the probability of accepting is at least $1-\frac{a}{2}$ if $H$ has an eigenvalue less than or equal to $am$, while this probability is at most $1-Cbn^{-c}$, for non-negative constants $c$ and $C$, in the case where $H$ has no eigenvalues below $bm$.

This leads directly to our first protocol for post hoc verification of a quantum computation $\mathcal{C}$:
\begin{enumerate}
\item The prover performs $\mathcal{C}$ obtaining an output string $S$ and sends this to the verifier.
\item The prover constructs a witness state $\ket{\eta}$ for the local Hamiltonian problem for $H_{\mathcal{V}_{(\mathcal{C},S)}^N}$, for suitably large $N$.
\item Once the verifier chooses to initiate verification, the prover encodes $\ket{\eta}$ in the 5-qubit code and distributes a share of each logical qubit among each of 5 spatially separated locations (effectively 5 provers) which can be queried in a space-like separated manner by the verifier.
\item The verifier then engages in the $5$-local Hamiltonian game described above. The verifier accepts $S$ as the output of the computation if and only if the provers win the game.
\end{enumerate}
From the completeness and soundness properties of the local Hamiltonian game, and from the bounds on the eigenvalues of $H_{\mathcal{V}_\mathcal{C}^S}$ imposed by the results of Kempe, Kitaev and Regev, it follows that the verifier will accept $S$ with polynomially higher probability when $p_S \geq 1-\delta$ than when $p_S \leq 1 - \delta -\gamma$ provided that $N > 2 \gamma^{-2} \log(C^{-1} n^c + 1)$, where $C$ and $c$ are the constants from the local Hamiltonian game in \cite{fitzsimons2015multiprover}.

Recently, Ji introduced a modification of the local Hamiltonian game considered above which removed the need for limited quantum computation on the part of the verifier \cite{ji2015classical}. This approach mirrors that of \cite{fitzsimons2015multiprover}, replacing the code-space and energy tests with versions that can be verified by a purely classical prover. 

The code-space test is accomplished using a clever application of CHSH rigidity. Consider the 5-qubit quantum error correction code with generators $\{g_i\}_{i=1}^4$ and let $\ket{\phi}$ be a state from the 2-dimensional stabilised subspace, that is $\bra{\phi}g_i\ket{\phi}=1$ for all $i$. The structure of the stabiliser generators is such that one of the subsystems, labelled $t$, always has either a Pauli $X$ or a Pauli $Z$ operator acting on it. Furthermore, due to translational invariance of the 5-qubit error correction code, we have freedom of choosing the subsystem $t$ and then fixing the remaining Pauli operators in $g_i$ appropriately while preserving the 2-dimensional code space. By a repeated use of a reflection operator $W_t=\cos\left(\frac{\pi}{8}\right)X_t+\sin\left(\frac{\pi}{8}\right)Z_t$ on the special subsystem $t$, we can obtain set of eight operators $\{h_i\}_{i=1}^8$ satisfying $\bra{\phi}\sum_{i=1}^8h_i\ket{\phi}=4\sqrt{2}$ and that are related to the original generators by $h_{2i-1}+h_{2i}=\sqrt{2}g_i$. Bipartitioning the 5 provers into non-special provers, labelled as system $A$, and the special prover $t$, labelled as $B$, we obtain the familiar CHSH expressions,
\begin{equation*}
    \bra{\phi} C \ket{\phi} = 2\sqrt{2}, \qquad \bra{\phi} C' \ket{\phi} = 2\sqrt{2},
\end{equation*}
where $C=\bar{X}_AH^+_B+\bar{X}_AH^-_B+\bar{Z}_AH^+_B-\bar{Z}_AH^-_B$ and $C'=\bar{X'}_AH^+_B+\bar{X'}_AH^-_B+\bar{Z'}_AH^+_B-\bar{Z'}_AH^-_B$. The Pauli operators $\bar{X}_A$ and $\bar{X'}_A$ are defined as the generators $g_i$ containing $X_t$ where we replace $X_t$ with the identity operator $I$. Similarly $\bar{Z}_A$ and $\bar{Z'}_A$ are defined as the generators $g_i$ whose $Z_t$ are replaced with $I$. For example when $t=5$, we have $\bar{X}_A=IXZZI$, $\bar{Z}_A=XIXZI$, $\bar{X'}_A=ZZXII$ and $\bar{Z'}_A=ZXIXI$. Via the rigidity of CHSH games, it can then be shown that if the probability of winning the corresponding CHSH game is close to ideal, then the shared state must be close to a state in the code-space.

The energy test is also replaced, making use of the fact that measurements of the logical $X$ and $Z$ operators of the code can be performed transversally by making an appropriate Pauli measurement on each share of the logical qubit. By expanding out the Hamiltonian as a sequence of Pauli terms polynomial in $n$. By choosing randomly to either measure a Pauli term from the Hamiltonian, or make Pauli measurements corresponding to a CHSH game to test the code space, it is possible to achieve an interactive proof for the local Hamiltonian problem where the verifier is entirely classical, with a polynomial gap between completeness and soundness, similar to that in \cite{fitzsimons2015multiprover}. 

It is thus possible to replace Step 4 in our initial protocol with a protocol that requires only a purely classical prover, using Ji's protocol. This removes the need for any quantum capabilities on the part of the prover while still retaining the post hoc verification characteristics of our first protocol, and requiring only one round of communication between verifier and provers. Recent modifications of this approach due to Natarajan and Vidick \cite{natarajan2015constant} could also be incorporated, potentially leading to improved performance.

\section*{Acknowledgements}

The authors acknowledge support from Singapore's National Research Foundation and Ministry of Education. JFF acknowldeges support from the Air Force Office of Scientific Research under grant FA2386-15-1-4082. This material is based on research funded by the Singapore National Research Foundation under NRF Award NRF-NRFF2013-01.

\bibliographystyle{apsrev}
\bibliography{final}

\begin{thebibliography}{36}
\expandafter\ifx\csname natexlab\endcsname\relax\def\natexlab#1{#1}\fi
\expandafter\ifx\csname bibnamefont\endcsname\relax
  \def\bibnamefont#1{#1}\fi
\expandafter\ifx\csname bibfnamefont\endcsname\relax
  \def\bibfnamefont#1{#1}\fi
\expandafter\ifx\csname citenamefont\endcsname\relax
  \def\citenamefont#1{#1}\fi
\expandafter\ifx\csname url\endcsname\relax
  \def\url#1{\texttt{#1}}\fi
\expandafter\ifx\csname urlprefix\endcsname\relax\def\urlprefix{URL }\fi
\providecommand{\bibinfo}[2]{#2}
\providecommand{\eprint}[2][]{\url{#2}}

\bibitem[{\citenamefont{Shor}(1994)}]{Shor1994}
\bibinfo{author}{\bibfnamefont{P.~W.} \bibnamefont{Shor}}, in
  \emph{\bibinfo{booktitle}{Foundations of Computer Science, 1994 Proceedings.,
  35th Annual Symposium on}} (\bibinfo{organization}{IEEE},
  \bibinfo{year}{1994}), pp. \bibinfo{pages}{124--134}.

\bibitem[{\citenamefont{Lloyd et~al.}(1996)}]{lloyd1996universal}
\bibinfo{author}{\bibfnamefont{S.}~\bibnamefont{Lloyd}} \bibnamefont{et~al.},
  \bibinfo{journal}{Science} pp. \bibinfo{pages}{1073--1077}
  (\bibinfo{year}{1996}).

\bibitem[{\citenamefont{Trotzky et~al.}(2012)\citenamefont{Trotzky, Chen,
  Flesch, McCulloch, Schollw{\"o}ck, Eisert, and Bloch}}]{trotzky2012probing}
\bibinfo{author}{\bibfnamefont{S.}~\bibnamefont{Trotzky}},
  \bibinfo{author}{\bibfnamefont{Y.-A.} \bibnamefont{Chen}},
  \bibinfo{author}{\bibfnamefont{A.}~\bibnamefont{Flesch}},
  \bibinfo{author}{\bibfnamefont{I.~P.} \bibnamefont{McCulloch}},
  \bibinfo{author}{\bibfnamefont{U.}~\bibnamefont{Schollw{\"o}ck}},
  \bibinfo{author}{\bibfnamefont{J.}~\bibnamefont{Eisert}}, \bibnamefont{and}
  \bibinfo{author}{\bibfnamefont{I.}~\bibnamefont{Bloch}},
  \bibinfo{journal}{Nature Physics} \textbf{\bibinfo{volume}{8}},
  \bibinfo{pages}{325} (\bibinfo{year}{2012}).

\bibitem[{\citenamefont{Boixo et~al.}(2013)\citenamefont{Boixo, Albash,
  Spedalieri, Chancellor, and Lidar}}]{boixo2013experimental}
\bibinfo{author}{\bibfnamefont{S.}~\bibnamefont{Boixo}},
  \bibinfo{author}{\bibfnamefont{T.}~\bibnamefont{Albash}},
  \bibinfo{author}{\bibfnamefont{F.~M.} \bibnamefont{Spedalieri}},
  \bibinfo{author}{\bibfnamefont{N.}~\bibnamefont{Chancellor}},
  \bibnamefont{and} \bibinfo{author}{\bibfnamefont{D.~A.} \bibnamefont{Lidar}},
  \bibinfo{journal}{Nature communications} \textbf{\bibinfo{volume}{4}}
  (\bibinfo{year}{2013}).

\bibitem[{\citenamefont{Smolin and Smith}(2013)}]{smolin2013classical}
\bibinfo{author}{\bibfnamefont{J.~A.} \bibnamefont{Smolin}} \bibnamefont{and}
  \bibinfo{author}{\bibfnamefont{G.}~\bibnamefont{Smith}},
  \bibinfo{journal}{arXiv preprint arXiv:1305.4904}  (\bibinfo{year}{2013}).

\bibitem[{\citenamefont{Wang et~al.}(2013)\citenamefont{Wang, R{\o}nnow, Boixo,
  Isakov, Wang, Wecker, Lidar, Martinis, and Troyer}}]{wang2013comment}
\bibinfo{author}{\bibfnamefont{L.}~\bibnamefont{Wang}},
  \bibinfo{author}{\bibfnamefont{T.~F.} \bibnamefont{R{\o}nnow}},
  \bibinfo{author}{\bibfnamefont{S.}~\bibnamefont{Boixo}},
  \bibinfo{author}{\bibfnamefont{S.~V.} \bibnamefont{Isakov}},
  \bibinfo{author}{\bibfnamefont{Z.}~\bibnamefont{Wang}},
  \bibinfo{author}{\bibfnamefont{D.}~\bibnamefont{Wecker}},
  \bibinfo{author}{\bibfnamefont{D.~A.} \bibnamefont{Lidar}},
  \bibinfo{author}{\bibfnamefont{J.~M.} \bibnamefont{Martinis}},
  \bibnamefont{and} \bibinfo{author}{\bibfnamefont{M.}~\bibnamefont{Troyer}},
  \bibinfo{journal}{arXiv preprint arXiv:1305.5837}  (\bibinfo{year}{2013}).

\bibitem[{\citenamefont{Boixo et~al.}(2014)\citenamefont{Boixo, R{\o}nnow,
  Isakov, Wang, Wecker, Lidar, Martinis, and Troyer}}]{boixo2014evidence}
\bibinfo{author}{\bibfnamefont{S.}~\bibnamefont{Boixo}},
  \bibinfo{author}{\bibfnamefont{T.~F.} \bibnamefont{R{\o}nnow}},
  \bibinfo{author}{\bibfnamefont{S.~V.} \bibnamefont{Isakov}},
  \bibinfo{author}{\bibfnamefont{Z.}~\bibnamefont{Wang}},
  \bibinfo{author}{\bibfnamefont{D.}~\bibnamefont{Wecker}},
  \bibinfo{author}{\bibfnamefont{D.~A.} \bibnamefont{Lidar}},
  \bibinfo{author}{\bibfnamefont{J.~M.} \bibnamefont{Martinis}},
  \bibnamefont{and} \bibinfo{author}{\bibfnamefont{M.}~\bibnamefont{Troyer}},
  \bibinfo{journal}{Nature Physics} \textbf{\bibinfo{volume}{10}},
  \bibinfo{pages}{218} (\bibinfo{year}{2014}).

\bibitem[{\citenamefont{R{\o}nnow et~al.}(2014)\citenamefont{R{\o}nnow, Wang,
  Job, Boixo, Isakov, Wecker, Martinis, Lidar, and
  Troyer}}]{ronnow2014defining}
\bibinfo{author}{\bibfnamefont{T.~F.} \bibnamefont{R{\o}nnow}},
  \bibinfo{author}{\bibfnamefont{Z.}~\bibnamefont{Wang}},
  \bibinfo{author}{\bibfnamefont{J.}~\bibnamefont{Job}},
  \bibinfo{author}{\bibfnamefont{S.}~\bibnamefont{Boixo}},
  \bibinfo{author}{\bibfnamefont{S.~V.} \bibnamefont{Isakov}},
  \bibinfo{author}{\bibfnamefont{D.}~\bibnamefont{Wecker}},
  \bibinfo{author}{\bibfnamefont{J.~M.} \bibnamefont{Martinis}},
  \bibinfo{author}{\bibfnamefont{D.~A.} \bibnamefont{Lidar}}, \bibnamefont{and}
  \bibinfo{author}{\bibfnamefont{M.}~\bibnamefont{Troyer}},
  \bibinfo{journal}{Science} \textbf{\bibinfo{volume}{345}},
  \bibinfo{pages}{420} (\bibinfo{year}{2014}).

\bibitem[{\citenamefont{Shin et~al.}(2014{\natexlab{a}})\citenamefont{Shin,
  Smith, Smolin, and Vazirani}}]{shin2014quantum}
\bibinfo{author}{\bibfnamefont{S.~W.} \bibnamefont{Shin}},
  \bibinfo{author}{\bibfnamefont{G.}~\bibnamefont{Smith}},
  \bibinfo{author}{\bibfnamefont{J.~A.} \bibnamefont{Smolin}},
  \bibnamefont{and} \bibinfo{author}{\bibfnamefont{U.}~\bibnamefont{Vazirani}},
  \bibinfo{journal}{arXiv preprint arXiv:1401.7087}
  (\bibinfo{year}{2014}{\natexlab{a}}).

\bibitem[{\citenamefont{Vinci et~al.}(2014)\citenamefont{Vinci, Albash, Mishra,
  Warburton, and Lidar}}]{vinci2014distinguishing}
\bibinfo{author}{\bibfnamefont{W.}~\bibnamefont{Vinci}},
  \bibinfo{author}{\bibfnamefont{T.}~\bibnamefont{Albash}},
  \bibinfo{author}{\bibfnamefont{A.}~\bibnamefont{Mishra}},
  \bibinfo{author}{\bibfnamefont{P.~A.} \bibnamefont{Warburton}},
  \bibnamefont{and} \bibinfo{author}{\bibfnamefont{D.~A.} \bibnamefont{Lidar}},
  \bibinfo{journal}{arXiv preprint arXiv:1403.4228}  (\bibinfo{year}{2014}).

\bibitem[{\citenamefont{Shin et~al.}(2014{\natexlab{b}})\citenamefont{Shin,
  Smith, Smolin, and Vazirani}}]{shin2014comment}
\bibinfo{author}{\bibfnamefont{S.~W.} \bibnamefont{Shin}},
  \bibinfo{author}{\bibfnamefont{G.}~\bibnamefont{Smith}},
  \bibinfo{author}{\bibfnamefont{J.~A.} \bibnamefont{Smolin}},
  \bibnamefont{and} \bibinfo{author}{\bibfnamefont{U.}~\bibnamefont{Vazirani}},
  \bibinfo{journal}{arXiv preprint arXiv:1404.6499}
  (\bibinfo{year}{2014}{\natexlab{b}}).

\bibitem[{\citenamefont{Albash et~al.}(2015)\citenamefont{Albash, R{\o}nnow,
  Troyer, and Lidar}}]{albash2015reexamining}
\bibinfo{author}{\bibfnamefont{T.}~\bibnamefont{Albash}},
  \bibinfo{author}{\bibfnamefont{T.~F.} \bibnamefont{R{\o}nnow}},
  \bibinfo{author}{\bibfnamefont{M.}~\bibnamefont{Troyer}}, \bibnamefont{and}
  \bibinfo{author}{\bibfnamefont{D.~A.} \bibnamefont{Lidar}},
  \bibinfo{journal}{The European Physical Journal Special Topics}
  \textbf{\bibinfo{volume}{224}}, \bibinfo{pages}{111} (\bibinfo{year}{2015}).

\bibitem[{\citenamefont{Aharonov et~al.}(2010)\citenamefont{Aharonov, Ben-Or,
  and Eban}}]{Aharonov:2008}
\bibinfo{author}{\bibfnamefont{D.}~\bibnamefont{Aharonov}},
  \bibinfo{author}{\bibfnamefont{M.}~\bibnamefont{Ben-Or}}, \bibnamefont{and}
  \bibinfo{author}{\bibfnamefont{E.}~\bibnamefont{Eban}}, in
  \emph{\bibinfo{booktitle}{Proceedings of Innovation in Computer Science}}
  (\bibinfo{publisher}{Tsinghua University Press}, \bibinfo{year}{2010}), p.
  \bibinfo{pages}{543}.

\bibitem[{\citenamefont{Fitzsimons and Kashefi}(2012)}]{Fitzsimons:2012}
\bibinfo{author}{\bibfnamefont{J.~F.} \bibnamefont{Fitzsimons}}
  \bibnamefont{and} \bibinfo{author}{\bibfnamefont{E.}~\bibnamefont{Kashefi}}
  (\bibinfo{year}{2012}), \eprint{quant-ph/1203.5217}.

\bibitem[{\citenamefont{Barz et~al.}(2013)\citenamefont{Barz, Fitzsimons,
  Kashefi, and Walther}}]{Barz:2013}
\bibinfo{author}{\bibfnamefont{S.}~\bibnamefont{Barz}},
  \bibinfo{author}{\bibfnamefont{J.~F.} \bibnamefont{Fitzsimons}},
  \bibinfo{author}{\bibfnamefont{E.}~\bibnamefont{Kashefi}}, \bibnamefont{and}
  \bibinfo{author}{\bibfnamefont{P.}~\bibnamefont{Walther}},
  \bibinfo{journal}{Nature Physics} \textbf{\bibinfo{volume}{9}},
  \bibinfo{pages}{727} (\bibinfo{year}{2013}).

\bibitem[{\citenamefont{Reichardt
  et~al.}(2013{\natexlab{a}})\citenamefont{Reichardt, Unger, and
  Vazirani}}]{Reichardt:2013}
\bibinfo{author}{\bibfnamefont{B.~W.} \bibnamefont{Reichardt}},
  \bibinfo{author}{\bibfnamefont{F.}~\bibnamefont{Unger}}, \bibnamefont{and}
  \bibinfo{author}{\bibfnamefont{U.}~\bibnamefont{Vazirani}},
  \bibinfo{journal}{Nature (London)} \textbf{\bibinfo{volume}{496}},
  \bibinfo{pages}{456} (\bibinfo{year}{2013}{\natexlab{a}}).

\bibitem[{\citenamefont{McKague}(2013)}]{Mckague:2013}
\bibinfo{author}{\bibfnamefont{M.}~\bibnamefont{McKague}}
  (\bibinfo{year}{2013}), \eprint{quant-ph/1309.5675}.

\bibitem[{\citenamefont{Morimae}(2014)}]{Morimae2014}
\bibinfo{author}{\bibfnamefont{T.}~\bibnamefont{Morimae}},
  \bibinfo{journal}{Phys. Rev. A} \textbf{\bibinfo{volume}{89}},
  \bibinfo{pages}{060302} (\bibinfo{year}{2014}).

\bibitem[{\citenamefont{Hayashi and Morimae}(2015)}]{hayashi2015verifiable}
\bibinfo{author}{\bibfnamefont{M.}~\bibnamefont{Hayashi}} \bibnamefont{and}
  \bibinfo{author}{\bibfnamefont{T.}~\bibnamefont{Morimae}},
  \bibinfo{journal}{arXiv preprint arXiv:1505.07535}  (\bibinfo{year}{2015}).

\bibitem[{\citenamefont{Reichardt
  et~al.}(2013{\natexlab{b}})\citenamefont{Reichardt, Unger, and
  Vazirani}}]{reichardt2013classical2}
\bibinfo{author}{\bibfnamefont{B.~W.} \bibnamefont{Reichardt}},
  \bibinfo{author}{\bibfnamefont{F.}~\bibnamefont{Unger}}, \bibnamefont{and}
  \bibinfo{author}{\bibfnamefont{U.}~\bibnamefont{Vazirani}}, in
  \emph{\bibinfo{booktitle}{Proceedings of the 4th conference on Innovations in
  Theoretical Computer Science}} (\bibinfo{organization}{ACM},
  \bibinfo{year}{2013}{\natexlab{b}}), pp. \bibinfo{pages}{321--322}.

\bibitem[{\citenamefont{Gheorghiu et~al.}(2015)\citenamefont{Gheorghiu,
  Kashefi, and Wallden}}]{gheorghiu2015robustness}
\bibinfo{author}{\bibfnamefont{A.}~\bibnamefont{Gheorghiu}},
  \bibinfo{author}{\bibfnamefont{E.}~\bibnamefont{Kashefi}}, \bibnamefont{and}
  \bibinfo{author}{\bibfnamefont{P.}~\bibnamefont{Wallden}},
  \bibinfo{journal}{arXiv preprint arXiv:1502.02571}  (\bibinfo{year}{2015}).

\bibitem[{\citenamefont{Hajdu{\v{s}}ek
  et~al.}(2015)\citenamefont{Hajdu{\v{s}}ek, P{\'e}rez-Delgado, and
  Fitzsimons}}]{hajduvsek2015device}
\bibinfo{author}{\bibfnamefont{M.}~\bibnamefont{Hajdu{\v{s}}ek}},
  \bibinfo{author}{\bibfnamefont{C.~A.} \bibnamefont{P{\'e}rez-Delgado}},
  \bibnamefont{and} \bibinfo{author}{\bibfnamefont{J.~F.}
  \bibnamefont{Fitzsimons}}, \bibinfo{journal}{arXiv preprint arXiv:1502.02563}
   (\bibinfo{year}{2015}).

\bibitem[{\citenamefont{Broadbent et~al.}(2009)\citenamefont{Broadbent,
  Fitzsimons, and Kashefi}}]{Broadbent:2009}
\bibinfo{author}{\bibfnamefont{A.}~\bibnamefont{Broadbent}},
  \bibinfo{author}{\bibfnamefont{J.~F.} \bibnamefont{Fitzsimons}},
  \bibnamefont{and} \bibinfo{author}{\bibfnamefont{E.}~\bibnamefont{Kashefi}},
  in \emph{\bibinfo{booktitle}{Foundations of Computer Science, 2009. FOCS'09.
  50th Annual IEEE Symposium on}} (\bibinfo{organization}{IEEE},
  \bibinfo{year}{2009}), pp. \bibinfo{pages}{517--526}.

\bibitem[{\citenamefont{Barz et~al.}(2012)\citenamefont{Barz, Kashefi,
  Broadbent, Fitzsimons, Zeilinger, and Walther}}]{Barz:2012}
\bibinfo{author}{\bibfnamefont{S.}~\bibnamefont{Barz}},
  \bibinfo{author}{\bibfnamefont{E.}~\bibnamefont{Kashefi}},
  \bibinfo{author}{\bibfnamefont{A.}~\bibnamefont{Broadbent}},
  \bibinfo{author}{\bibfnamefont{J.~F.} \bibnamefont{Fitzsimons}},
  \bibinfo{author}{\bibfnamefont{A.}~\bibnamefont{Zeilinger}},
  \bibnamefont{and} \bibinfo{author}{\bibfnamefont{P.}~\bibnamefont{Walther}},
  \bibinfo{journal}{Science} \textbf{\bibinfo{volume}{335}},
  \bibinfo{pages}{303} (\bibinfo{year}{2012}).

\bibitem[{\citenamefont{Fitzsimons and
  Vidick}(2015)}]{fitzsimons2015multiprover}
\bibinfo{author}{\bibfnamefont{J.}~\bibnamefont{Fitzsimons}} \bibnamefont{and}
  \bibinfo{author}{\bibfnamefont{T.}~\bibnamefont{Vidick}}, in
  \emph{\bibinfo{booktitle}{Proceedings of the 2015 Conference on Innovations
  in Theoretical Computer Science}} (\bibinfo{organization}{ACM},
  \bibinfo{year}{2015}), pp. \bibinfo{pages}{103--112}.

\bibitem[{\citenamefont{Ji}(2015)}]{ji2015classical}
\bibinfo{author}{\bibfnamefont{Z.}~\bibnamefont{Ji}}, \bibinfo{journal}{arXiv
  preprint arXiv:1505.07432}  (\bibinfo{year}{2015}).

\bibitem[{\citenamefont{Kitaev}(1999)}]{kitaev1999quantum}
\bibinfo{author}{\bibfnamefont{A.}~\bibnamefont{Kitaev}},
  \bibinfo{journal}{Talk at AQIP} \textbf{\bibinfo{volume}{99}}
  (\bibinfo{year}{1999}).

\bibitem[{\citenamefont{Aharonov and Naveh}(2002)}]{aharonov2002quantum}
\bibinfo{author}{\bibfnamefont{D.}~\bibnamefont{Aharonov}} \bibnamefont{and}
  \bibinfo{author}{\bibfnamefont{T.}~\bibnamefont{Naveh}},
  \bibinfo{journal}{arXiv preprint quant-ph/0210077}  (\bibinfo{year}{2002}).

\bibitem[{\citenamefont{Kempe and Regev}(2003)}]{kempe20033}
\bibinfo{author}{\bibfnamefont{J.}~\bibnamefont{Kempe}} \bibnamefont{and}
  \bibinfo{author}{\bibfnamefont{O.}~\bibnamefont{Regev}},
  \bibinfo{journal}{Quantum Information \& Computation}
  \textbf{\bibinfo{volume}{3}}, \bibinfo{pages}{258} (\bibinfo{year}{2003}).

\bibitem[{\citenamefont{Kempe et~al.}(2006{\natexlab{a}})\citenamefont{Kempe,
  Kitaev, and Regev}}]{kempe2006complexity}
\bibinfo{author}{\bibfnamefont{J.}~\bibnamefont{Kempe}},
  \bibinfo{author}{\bibfnamefont{A.}~\bibnamefont{Kitaev}}, \bibnamefont{and}
  \bibinfo{author}{\bibfnamefont{O.}~\bibnamefont{Regev}},
  \bibinfo{journal}{SIAM Journal on Computing} \textbf{\bibinfo{volume}{35}},
  \bibinfo{pages}{1070} (\bibinfo{year}{2006}{\natexlab{a}}).

\bibitem[{\citenamefont{Dunjko et~al.}(2014)\citenamefont{Dunjko, Fitzsimons,
  Portmann, and Renner}}]{Dunjko2014}
\bibinfo{author}{\bibfnamefont{V.}~\bibnamefont{Dunjko}},
  \bibinfo{author}{\bibfnamefont{J.~F.} \bibnamefont{Fitzsimons}},
  \bibinfo{author}{\bibfnamefont{C.}~\bibnamefont{Portmann}}, \bibnamefont{and}
  \bibinfo{author}{\bibfnamefont{R.}~\bibnamefont{Renner}}, in
  \emph{\bibinfo{booktitle}{Advances in Cryptology--ASIACRYPT 2014}}
  (\bibinfo{publisher}{Springer}, \bibinfo{year}{2014}), pp.
  \bibinfo{pages}{406--425}.

\bibitem[{\citenamefont{Aaronson}(2005)}]{postselection}
\bibinfo{author}{\bibfnamefont{S.}~\bibnamefont{Aaronson}}, in
  \emph{\bibinfo{booktitle}{Proceedings of the Royal Society of London A:
  Mathematical, Physical and Engineering Sciences}} (\bibinfo{organization}{The
  Royal Society}, \bibinfo{year}{2005}), vol. \bibinfo{volume}{461}, pp.
  \bibinfo{pages}{3473--3482}.

\bibitem[{\citenamefont{Hoeffding}(1963)}]{hoeffding1963probability}
\bibinfo{author}{\bibfnamefont{W.}~\bibnamefont{Hoeffding}},
  \bibinfo{journal}{Journal of the American statistical association}
  \textbf{\bibinfo{volume}{58}}, \bibinfo{pages}{13} (\bibinfo{year}{1963}).

\bibitem[{\citenamefont{Kempe et~al.}(2006{\natexlab{b}})\citenamefont{Kempe,
  Kitaev, and Regev}}]{KempeKitaevRegev}
\bibinfo{author}{\bibfnamefont{J.}~\bibnamefont{Kempe}},
  \bibinfo{author}{\bibfnamefont{A.}~\bibnamefont{Kitaev}}, \bibnamefont{and}
  \bibinfo{author}{\bibfnamefont{O.}~\bibnamefont{Regev}},
  \bibinfo{journal}{SIAM Journal on Computing} \textbf{\bibinfo{volume}{35}},
  \bibinfo{pages}{1070} (\bibinfo{year}{2006}{\natexlab{b}}).

\bibitem[{\citenamefont{Natarajan and Vidick}(2015)}]{natarajan2015constant}
\bibinfo{author}{\bibfnamefont{A.}~\bibnamefont{Natarajan}} \bibnamefont{and}
  \bibinfo{author}{\bibfnamefont{T.}~\bibnamefont{Vidick}},
  \bibinfo{journal}{arXiv preprint arXiv:1512.02090}  (\bibinfo{year}{2015}).

\bibitem[{\citenamefont{Morimae and Fujii}(2013)}]{Morimae:2013}
\bibinfo{author}{\bibfnamefont{T.}~\bibnamefont{Morimae}} \bibnamefont{and}
  \bibinfo{author}{\bibfnamefont{K.}~\bibnamefont{Fujii}},
  \bibinfo{journal}{Phys. Rev. A} \textbf{\bibinfo{volume}{87}},
  \bibinfo{pages}{050301} (\bibinfo{year}{2013}).

\end{thebibliography}
\end{document}